\documentclass[aps,prb,twocolumn,superscriptaddress,floatfix,longbibliography,10pt]{revtex4-2}

\usepackage{amsmath,amssymb,mathtools,bm}
\usepackage{graphicx}
\usepackage{physics}
\usepackage{siunitx}
\usepackage{csquotes}
\usepackage{longtable}
\usepackage{float}

\usepackage{hyperref}

\graphicspath{{./IMG/}}

\begin{document}

\title{Machine Learning-Driven High-Throughput Screening of Equimolar and Non-Equimolar High-Entropy Carbides}

\author{Arina A. Svinukhova$^+$} 
\affiliation{Tomsk Polytechnic University, Tomsk, Russian Federation}

\author{Faisal Baber$^+$}
\affiliation{National Research University Higher School of Economics (HSE University), Russian Federation}

\author{Michel M. Lukanov} 
\affiliation{Skolkovo Institute of Science and Technology, Bolshoy Boulevard 30, bld. 1, Moscow 121205, Russian Federation}

\author{Viktor S. Baidyshev}
\affiliation{Skolkovo Institute of Science and Technology, Bolshoy Boulevard 30, bld. 1, Moscow 121205, Russian Federation}

\author{Daria Terentieva}
\affiliation{Tomsk Polytechnic University, Tomsk, Russian Federation}

\author{Alexander Ya. Pak}
\email{ayapak@tpu.ru}
\affiliation{Tomsk Polytechnic University, Tomsk, Russian Federation}

\author{Alexander G. Kvashnin}
\email{A.Kvashnin@skoltech.ru}
\affiliation{Skolkovo Institute of Science and Technology, Bolshoy Boulevard 30, bld. 1, Moscow 121205, Russian Federation}

\begin{abstract}
Search for optimal compositions among high-entropy ceramics are challenging due to large composition space, which are unable to test either experimentally or by \textit{ab initio} calculations.
Here we present a high-throughput screening for equimolar and non-equimolar high-entropy carbides (HEC) among group IVB and VB transition metals (Ti, Zr, Hf, V, Nb and Ta). 
Large Atomistic Model MACE, and deep learning based interatomic potential DeePMD trained on \textit{ab initio}-based dataset of structures are used to effectively screen over the thousands of potential non-equimolar compositions among high-entropy carbides of Ti, Zr, Hf, V, Nb and Ta transition metals. 
A set of both equimolar and non-equimolar HECs are predicted to be stable in order to further synthesize those compositions based on the predictions. 
\end{abstract}

\maketitle

\section{Introduction}
In recent years, there has been a growing interest in developing new materials with properties that meet the growing needs for strength, durability, and resistance to aggressive operating conditions.
One of the promising areas is the production of high-entropy materials.
The first mention of materials with high entropy was associated with the production of high-entropy alloys in 2004 \cite{yeh2004nanostructured}.
Subsequently, a number of high-entropy materials expanded. 
Efforts have been made to obtain and to study the properties of high-entropy borides\cite{gild2016highentropy} , oxides \cite{rost2015entropystabilized}, carbides \cite{sun2021singlephase,demirskyi2020synthesis}, nitrides \cite{jin2018mechanochemical}, sulfides \cite{zhang2018datadriven} and silicides \cite{qin2019highentropy}.
Today, materials capable of withstanding high and ultra-high temperatures are in high demand for use both in the aerospace industry and in the energy sector \cite{gomez-martin2016permeability}.
A potential candidate for use in such conditions is ultrahigh temperature ceramics (UHTC), which is characterized by a melting point of 3300 K \cite{gomez-martin2016permeability}. 
However, the currently existing representatives of high-temperature ceramics are not always able to meet the growing needs in some aspects of the industry, which leads to the research of new materials, in particular, high-entropy carbides (HEC) \cite{li2021phase}.

High-entropy carbide is a solid solution of four or more principal elements of transition metal carbides in almost equiatomic ratios or at least with a content of 5 to 35 at.\% in a single-phase structure \cite{li2022synthesis}. 
High-entropy ceramics attract a lot of attention due to their unique properties and wide range of potential applications.
It was found that this class of materials has high melting point, hardness and strength, low thermal conductivity, corrosion and radiation resistance, and also has good catalytic properties \cite{haglund1993theory,oyama1996chemistry,castle2018processing,harrington2019phase,yan2018hfzrtanbti,tunes2023from,niu2022electrical}.
Thus, high-entropy carbides are excellent candidates for use as structural components in the aerospace, metallurgical, mining, and nuclear industries \cite{chicardi2019low,wei2020gradient}.

The large variability of high-entropy carbides makes it possible to synthesize materials with various unique properties, but makes experimental research on their search rather difficult. 
In this regard, attempts have been made to theoretically describe the characteristics inherent in high-entropy carbides, which could be used to predict the potential of obtaining their single-phase combinations and their physico-mechanical properties \cite{kaufmann2020discovery,mellor2021development,pak2023machine,sarker2018highentropy}.
Thus, work has already been carried out on calculations of fifteen high-entropy equiatomic quaternary carbides of refractory metals of groups IVB and VB (Ti, Zr, Hf, V, Nb and Ta) and their mechanical properties.
As a result, it was found out that all 15 variations satisfy the criterion of formation and can be obtained experimentally, which makes it meaningful to further attempts to synthesize previously unknown combinations \cite{liu2021phase}.
In a recently published paper by Hong Meng et al. \cite{meng2023formation} 91 HEC samples from metals of the IVB, VB, and VIB groups (Ti, Zr, Hf, V, Nb, Ta, Cr, Mo, and W) were synthesized using high-performance ultrafast high-temperature synthesis, and 22 shape descriptors were simultaneously calculated using high-performance calculations. 
Based on the obtained results \cite{meng2023formation}, eight key descriptors characterizing the ability to form new structures were identified using machine learning and a genetic algorithm.

In addition to predicting the possibility of obtaining certain variations of high-entropy carbides, a synthesis method is needed that allows one to experimentally validate the theoretical calculations.
It is important that the technique should be effective, easy to use, does not require large economic costs for conducting experiments, and allows achieving the temperatures necessary for the synthesis of high-entropy carbides. 
One of the potential techniques is vacuum-free electric arc synthesis.
This method implements the synthesis of materials in an open air environment during the combustion of an arc discharge, which is possible as a result of the effect of self-shielding of the reaction zone by released gases of monoxide and carbon dioxide \cite{pak2020vacuumless,pak2022novel}.
Due to the implementation of the process in an air environment, there is no need to use a sealed chamber, vacuum equipment and other additional systems to provide an inert environment \cite{su2014lowcost}.
Previously, this technique has already been successfully tested to obtain TiZrNbHfTaC$_5$ \cite{pak2023machine}.

This paper presents the results of a study of theoretical calculations of the possibility of the existence of equiatomic single-phase high-entropy carbides consisting of four metals of IVb and Vb groups (Ti, Zr, Hf, V, Nb and Ta).

\section{Computational Methods}

\subsection{Construction of HEC dataset}

To construct the HEC dataset we consider six metals (Ti, Zr, Hf, V, Nb and Ta) occupying cation position in the rocksalt structure with unit cell containing 64 atoms (32 cations, 32 anions). 
Only equimolar concentrations of each principal component are considered for initial dataset.
All principal elements across the unit cell volume are randomly mixed in order to create different local arrangements of each atoms.
In total a dataset of 3050 HEC configurations is created.

All total-energy calculations and structural optimizations of each generated HEC structure are performed using the Vienna \emph{ab initio} Simulation Package (VASP).\cite{Kresse1996VASP,Hafner2008VASP}
Spin-polarized density functional theory (DFT) is employed throughout within the projector augmented-wave (PAW) formalism.\cite{Blochl1994PAW,Kresse1999PAW}
Exchange--correlation effects are described using the revised Perdew--Burke--Ernzerhof (revPBE) generalized gradient approximation,\cite{Hammer1999revPBE} which is selected because it generally provides a more balanced description of structure and properties of carbide systems than standard GGA functionals \cite{baidyshev2024melting}.

The Kohn--Sham orbitals are expanded in a plane-wave basis set with a kinetic-energy cutoff of 480~eV.
Electronic occupations are treated using Gaussian smearing with a width of 0.05~eV.
During ionic optimization, the electronic self-consistency threshold at each step is set to $10^{-5}$~eV.

\subsection{Training of MLIPs on HEC dataset}

\subsubsection{MACE}
The first interatomic potential trained on the HEC dataset is based on the MACE architecture, a higher order equivariant message passing neural network that has demonstrated strong transferability across multi-component metallic and ceramic systems.
Rather than training a MACE model from randomly initialized weights, the foundation model MACE-MPA-0 is used as the starting point, and all of its weights are allowed to vary during fine-tuning. This foundation model was selected for consistency with the approach taken by Magorrian et al. \cite{magorrian2025computationaltuningelasticproperties}, who fine-tuned the same foundation model for the related task of predicting elastic properties of low- and high-entropy ultra-high temperature ceramics. 
The MACE-MPA-0 architecture comprises two interaction layers with 128 invariant and equivariant feature channels, a maximum equivariance order of $L=1$, a correlation order of $\nu=3$, and a cutoff radius of 6~\AA\ for the local atomic environment.

The HEC dataset of 3050 configurations is partitioned into training, validation, and internal test sets in proportions of 2412, 286, and 353 configurations, respectively.
A force-based filter is subsequently applied to the training set, removing one configuration whose maximum atomic force exceeds 20~eV/\AA, leaving 2411 configurations used for fine-tuning. 
Energies and forces are read from the dataset using the REF\_energy and REF\_forces keys to ensure compatibility with the MACE training utilities.

Fine-tuning is carried out with a batch size of 1 for a maximum of 300 epochs, using the AMSGrad variant of the Adam optimizer with an exponential moving average of weight decay 0.99.
During the first 200 epochs, the loss function combines energy and force terms with a weight ratio of 1:100 and a learning rate of 0.01.
Stochastic weight averaging is then activated at epoch 200, at which point the learning rate is reduced to $10^{-4}$ and the energy-to-force weight ratio is increased to 1000:100, favoring tighter convergence of the energy predictions in the later stages of training. 
The final model, designated MACE-UHTC-HEC, is taken from the stochastic weight averaging checkpoint at epoch 222, selected as the best-performing checkpoint among those produced after the onset of weight averaging.

On the internal test set, MACE-UHTC-HEC achieves an energy mean absolute error of 1.70~meV/atom and a force mean absolute error of 38.0~meV/\AA.
The energy accuracy surpasses the 2.5~meV/atom benchmark adopted for this dataset, despite the force error remaining above the corresponding benchmark of 19.6~meV/\AA, reflecting the broader compositional diversity of the present seven-element chemical space relative to the more restricted systems considered in earlier work.
When evaluated on an independent blind test set of 153 frames not used at any stage of training, validation, or model selection, the fine-tuned model attains an even lower energy mean absolute error of 1.40 ~meV/atom and a force mean absolute error of 21.6~meV/\AA, confirming that the model generalizes well beyond the configurations seen during fine-tuning.

\subsubsection{DeePMD}
A second interatomic potential is trained on the same HEC dataset using the DeePMD-kit framework (v3.1.3, PyTorch backend)\cite{wang2018deepmd,zeng2023deepmd}, allowing for a direct comparison against a model built entirely from scratch rather than fine-tuned from a pre-trained foundation model.
The descriptor is of the smooth edition type se\_e2\_a, with a cutoff radius of 6~\AA\ and a smoothing radius of 5.5~\AA.
The neighbor selection list is set to 30 atoms per species across all seven elements of the chemical space (Hf, Nb, Ta, Ti, V, Zr, C).
The embedding network consists of three layers of 25, 50, and 100 neurons without a residual connection on the time step, followed by an axis neuron dimension of 16.
The fitting network maps the resulting descriptor onto the atomic energy through three hidden layers of 240 neurons each, with a residual connection on the time step enabled.

The model is trained using the same training and validation partitions of the HEC dataset employed for MACE-UHTC-HEC. 
The loss function combines an energy term and a force term, with the energy prefactor increasing from a starting value of 0.02 to a limiting value of 1, and the force prefactor decreasing from a starting value of 1000 to a limiting value of 1, following the standard DeePMD-kit energy loss schedule.
The learning rate follows an exponential decay schedule, starting at $10^{-3}$ and decaying with a decay step of 5000, down to a final value of $3.51\times10^{-8}$. Training is carried out for 500{,}000 steps on a single Tesla V100-SXM2-32GB GPU, with checkpoints saved every 50{,}000 steps. The resulting model is frozen and stored as deepmd\_model.pth.

On the same internal test set used to evaluate MACE-UHTC-HEC, the DeePMD model achieves an energy mean absolute error of 4.06~meV/atom and a force mean absolute error of 84.3~meV/\AA. 
These errors are approximately 2.4 times and 2.2 times larger, respectively, than those obtained with the fine-tuned MACE-UHTC-HEC model.

\begin{table}[H]
\centering
\caption{Comparison of training configuration and accuracy on the internal test set for the fine-tuned MACE-UHTC-HEC model and the DeePMD model trained from scratch.}
\label{tab:mace_vs_deepmd}
\resizebox{\columnwidth}{!}{%
\begin{tabular}{lcc}
\hline
 & MACE-UHTC-HEC & DeePMD \\
\hline
Initialization        & MACE-MPA-0 foundation model & Random (from scratch) \\
Cutoff radius          & 6~\AA                        & 6~\AA \\
Batch size             & 1                            & 10 \\
Max.\ epochs / steps   & 300 epochs                   & 500{,}000 steps \\
Energy MAE (test)      & 1.69~meV/atom                & 4.06~meV/atom \\
Force MAE (test)       & 38.0~meV/\AA                 & 84.3~meV/\AA \\
\hline
\end{tabular}%
}
\end{table}

\subsection{PCA and t-SNE visualization}
Dimensionality reduction was performed using scikit-learn (v1.5.2). 
Each screened composition was represented as an 8-dimensional feature vector: the six metal site fractions ($x_{\mathrm{Hf}}, x_{\mathrm{Nb}}, x_{\mathrm{Ta}}, x_{\mathrm{Ti}}, x_{\mathrm{V}}, x_{\mathrm{Zr}}$), the
mixing energy per atom $E_{\mathrm{mix}}^{\mathrm{atom}}$, and the
effective stabilization temperature $T^*$.
Features were standardized to zero mean and unit variance (\texttt{StandardScaler}) prior to reduction.
Principal component analysis (\texttt{sklearn.decomposition.PCA}, 2 components, \texttt{random\_state=42}) was applied to the full combined set of 7{,}636 points -- comprising all 7{,}581 raw non-equimolar screening points (including collapsed-subset duplicates identified in Sec.~\ref{sec:nonequimolar}) together with the 55 equimolar-ensemble compositions -- with the first two components explaining 38.7\% and 15.0\% of the variance, respectively. 
For t-SNE (\texttt{sklearn.manifold.TSNE}), a random subsample of 3{,}000 points was used for computational tractability, with perplexity 40, learning rate set to \texttt{`auto'}, PCA initialization, 1{,}000 iterations, and \texttt{random\_state=42}.

\subsubsection{Descriptor for synthesizability, $T^*$}
A central quantity used to rank the synthesizability of candidate high-entropy carbide compositions is the effective stabilization temperature, $T^*$, following the formalism introduced by Magorrian et al. \cite{magorrian2025computationaltuningelasticproperties}. 
High-entropy compounds are stabilized against phase segregation by the configurational disorder of their constituent elements on the metallic sublattice.
This mixing entropy contributes a $-TS$ term to the free energy that grows with temperature, while the mixing itself typically carries an energy penalty relative to the unmixed binary carbides.

Thus, the effective stabilization temperature is defined as the temperature at which these two contributions exactly cancel,

\begin{equation}
T^* = \frac{E^{\text{atom}}_{\text{mix}}}{S^{\text{atom}}_{\text{conf}}},
\label{eq:Tstar}
\end{equation}

where $S^{\text{atom}}_{\text{conf}}$ is the configurational mixing entropy per atom. 
This is obtained from the mixing entropy per metallic site,

\begin{equation}
S^{\text{site}}_{\text{conf}} = -k_B \sum_i x_i \ln x_i,
\label{eq:Sconf}
\end{equation}

where $\{x_i\}$ are the fractional occupancies of the metal species on the metallic sublattice, normalized by the fraction of metal atoms in the cell, $N_{\text{metal}}/N_{\text{tot}} = 0.5$ for the rock-salt carbide structures considered here, to give $S^{\text{atom}}_{\text{conf}} = (N_{\text{metal}}/N_{\text{tot}})\, S^{\text{site}}_{\text{conf}}$.

The mixing energy per atom, $E^{\text{atom}}_{\text{mix}}$, is computed as the difference between the energy per atom of the relaxed mixture configuration and the composition-weighted sum of the reference energies of the corresponding single-metal binary carbides,

\begin{equation}
E^{\text{atom}}_{\text{mix}} = E^{\text{atom}}_{\text{conf}} - \sum_i x_i E_i,
\label{eq:Emix}
\end{equation}

with both $E^{\text{atom}}_{\text{conf}}$ and the binary reference energies $E_i$ evaluated using the fine-tuned MACE-UHTC-HEC model.
As $T^*$ does not include vibrational or electronic free energy contributions, it should not be interpreted as a literal synthesis temperature; rather, it serves as a computationally inexpensive descriptor for ranking compositions by their relative ease of stabilization.
Negative values of $T^*$ indicate that a composition is already energetically favorable at zero temperature, without any contribution from the mixing entropy, whereas larger positive values indicate a composition that requires a correspondingly larger thermal entropic contribution to become stable.

Compositions are classified into four bands according to the value of $T^*$: stable at $T=0$~K for $T^*<0$~K, synthesizable for $0\leq T^*<1500$~K, difficult for $1500\leq T^*<3000$~K, and likely unstable for $T^*\geq3000$~K. This classification is applied uniformly across both the equimolar and non-equimolar composition screening described in subsequent sections.

\section{Results and discussion}

\subsection{Distribution of the data across the dataset}\label{sec:dataset_distribution}
The dataset used for training and evaluating the interatomic potentials comprises 3050 DFT-derived configurations spanning 55 distinct chemical compositions, drawn from the full seven-element space of Hf, Nb, Ta, Ti, V, Zr, and C. 
Configurations are built from supercells containing either 64 or 128 atoms, with each metal sublattice site occupied according to the metal fractions defining the corresponding composition. 
The energy per atom across the dataset ranges from $-10.3297$ to $-9.0832$~eV, while the per-atom force magnitudes have a mean value of 1.21~eV/\AA\ and a maximum value of 79.0~eV/\AA, reflecting the inclusion of randomly strained and rattled configurations alongside relaxed structures.

The number of different configurations per composition is highly non-uniform. 
The equimolar six-metal composition Hf$_5$Nb$_5$Ta$_6$Ti$_5$V$_6$Zr$_5$C$_{32}$ dominates the dataset with 1030 configurations, an order of magnitude larger than the next most represented compositions, which include the quaternary Nb$_8$Ta$_8$Ti$_8$V$_8$C$_{32}$ with 119 configurations and the quaternary Hf$_8$Ta$_8$V$_8$Zr$_8$C$_{32}$ with 86 configurations.
Of the 55 compositions, 47 are classified as high confidence, containing at least 20 configurations each, while a smaller number of compositions fall below the low-confidence threshold of 5 configurations.
This imbalance reflects the emphasis placed during dataset construction on densely sampling the fully equimolar six-metal carbide, which represents the primary target composition of the wider screening effort, while still providing broad coverage of lower-order binary, ternary, and quaternary carbides across the same chemical space.

To ensure that this imbalance does not bias the assessment of model performance on any single composition, the dataset is partitioned into training, validation, and internal test sets using a stratified split applied independently within each composition, rather than a single random split applied to the dataset as a whole. 
For each composition, HEC configurations are partitioned in an approximate 80/10/10 ratio, with compositions containing fewer than three configurations placed entirely into the training set.
This stratified scheme results in a total split of 2412 training, 286 validation, and 353 test structural configurations, ensuring that every composition with sufficient number of structures is represented in the validation and test sets and that the reported model accuracy is not dominated by a single dominant composition.

\subsection{Validation of MLIPs}
For calculations of energies of considered HECs two interatomic potentials, namely the fine-tuned MACE-UHTC-HEC and the DeePMD models are trained and validated on the internal test set of 353 configurations described in Sec.~\ref{sec:dataset_distribution}, which is withheld from training and validation for both models.
This internal test set spans the full range of compositions present in the dataset, including the dominant equimolar six-metal carbide as well as the lower-order binary, ternary, and quaternary carbides, providing a representative assessment of each model across the chemical space rather than a single composition.

\begin{figure}[ht]
\centering
\includegraphics[width=\columnwidth]{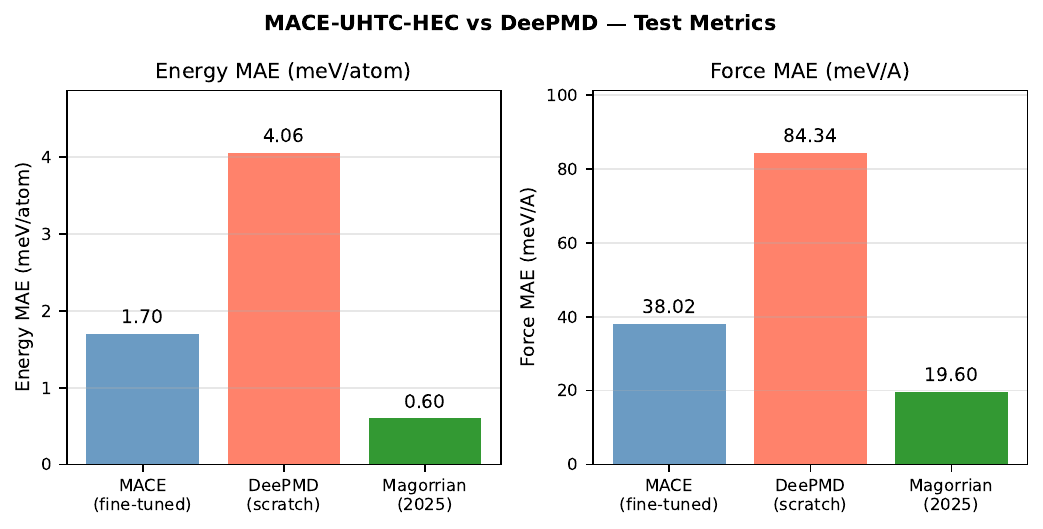}
\caption{Comparison of energy and force mean absolute errors on the internal test set of 353 configurations for the fine-tuned MACE-UHTC-HEC model and the DeePMD model trained from scratch, alongside the benchmark values reported by Magorrian et al. \cite{magorrian2025computationaltuningelasticproperties}.}
\label{fig:step2_mace_vs_deepmd}
\end{figure}

On this test set, MACE-UHTC-HEC achieves an energy mean absolute error of 1.70~meV/atom and a force mean absolute error of 38.0~meV/\AA, while the DeePMD model achieves an energy mean absolute error of 4.06~meV/atom and a force mean absolute error of 84.3~meV/\AA.
Figure~\ref{fig:step2_mace_vs_deepmd} summarizes this comparison alongside the corresponding benchmark values reported by Magorrian et al. \cite{magorrian2025computationaltuningelasticproperties} for a related fine-tuned MACE model evaluated on a smaller chemical space.
The fine-tuned MACE-UHTC-HEC model reduces the energy error by a factor of $\sim$2.4 and the force error by a factor of $\sim$2.2 relative to the DeePMD model trained from scratch on the same data, despite both models being trained on identical training and validation partitions.
This indicates that the prior physical and chemical knowledge encoded in the MACE-MPA-0 foundation model transfers effectively to the seven-element UHTC-HEC chemical space, providing a substantial accuracy advantage over a comparable architecture trained without this prior knowledge.

Given its substantially lower error on both energies and forces, MACE-UHTC-HEC is adopted as the primary interatomic potential for all subsequent high-throughput screening, blind testing, and compositional analysis presented in the following sections, with the DeePMD model retained solely as a comparative baseline.

\subsection{Blind test of MLIPs on non-equimolar HECs}\label{sec:blind_noneq}
To further assess the generalization capability of the fine-tuned MACE-UHTC-HEC model beyond the internal test set, the model is additionally evaluated on an independent blind test set of 153 configurations.
We have generated 153 non-equimolar HECs containing the same principal elements (Ti, Zr, Hf, V, Nb and Ta) with the variation of the concentration of each metal atom from 0 to 100\% with an 10\% increment.
All generated 153 configurations are relaxed using DTF within the same accuracy parameters used for entire dataset for training. 
This blind test set is not used at any stage of training, validation, or model and checkpoint selection, and is therefore used solely as an external check on model performance after the final SWA checkpoint at epoch 222 had already been selected based on the internal validation and test sets described in Section~\ref{sec:dataset_distribution}. 

\begin{figure}[ht]
\centering
\includegraphics[width=\columnwidth]{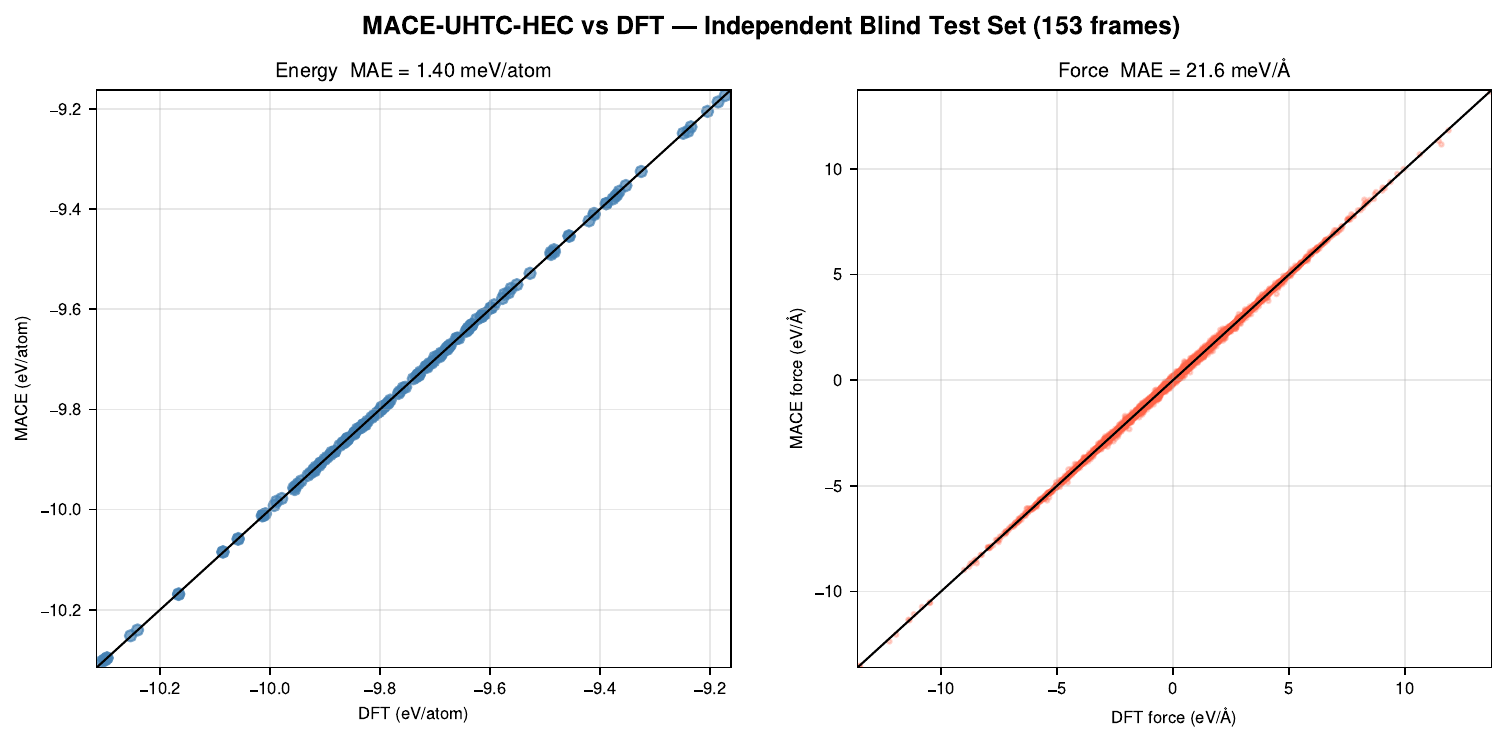}
\caption{Parity plots comparing MACE-UHTC-HEC predictions against DFT reference values for energy and force components on the independent 153-configuration blind test set of non-equimolar HECs.}
\label{fig:blind_test_parity}
\end{figure}

On this blind test set, MACE-UHTC-HEC achieves an energy mean absolute error of 1.40 ~meV/atom and a force mean absolute error of 21.6~meV/\AA. 
Figure~\ref{fig:blind_test_parity} shows the resulting parity plots for energy and force components, comparing the model predictions directly against the DFT reference values across all 153 blind configurations of non-equimolar HECs.
Both the energy and force predictions lie tightly along the parity line, with no systematic deviation visible across the full range of energies sampled, from approximately $-10.3$ to $-9.2$~eV/atom, or across the full range of force components shown.

Notably, the blind test errors are slightly lower than the corresponding errors obtained on the internal test set, with the energy mean absolute error decreasing from 1.70 to 1.40 ~meV/atom and the force mean absolute error decreasing from 38.0 to 21.6~meV/\AA. 
This indicates that the model accuracy reported on the internal test set is not an artifact of any residual correlation between the training and test partitions arising from the stratified per-composition split, and confirms that MACE-UHTC-HEC generalizes reliably to configurations entirely outside of the dataset used for its development.

\subsection{Generation of non-equimolar HECs}
\label{sec:nonequimolar}
As the blind test on non-equimolar HEC configurations is successfully passed (see Section \ref{sec:blind_noneq}), we move to high-throughput screening among full-composition configuration space of  non-equimolar HECs.   
Beyond the dense sampling of equimolar mixture compounds described in the preceding sections, the fine-tuned MACE-UHTC-HEC model is used to screen non-equimolar compositions across the full combinatorial space of metal subsets. 
All 57 non-empty subsets of the six transition metals Hf, Nb, Ta, Ti, V, and Zr are considered, comprising 15 binary, 20 ternary, 15 quaternary, 6 quinary, and 1 senary subset. 
For each subset, the corresponding metal fractions are sampled on a regular composition grid in increments of one eighth, generating a dense set of non-equimolar compositions within each subset while keeping the screening computationally tractable.

For every sampled composition, a supercell is constructed with the corresponding metal occupancies, relaxed using the fine-tuned MACE-UHTC-HEC model, and the resulting relaxed energy is used to compute the mixing energy per atom and the effective stabilization temperature $T^*$ as defined in Eq.~\eqref{eq:Tstar}. 
This procedure yields a total of 7581 composition points across the 57 subsets, with 105 points for the binary subsets, 840 for the ternary subsets, 2415 for the quaternary subsets, 2940 for the quinary subsets, and 1281 for the senary subset.

A degeneracy is identified in this screening procedure, arising from the composition grid occasionally placing one or more metal fractions below the threshold used to determine whether a metal is genuinely active in a given configuration. 
When this occurs, a composition nominally belonging to a higher-order subset collapses onto a lower-order subset that has already been sampled independently, producing a duplicate entry rather than a genuine non-equimolar point. 
Of the 7581 total screening points, 6300, or 83.1\%, are found to be such collapsed-subset rediscoveries, leaving 1281 genuine non-equimolar points.
This is corrected by applying a filter requiring that the number of metals with a fraction above the activity threshold, $n_{\text{active}}$, equals the nominal number of metals defining the subset, $n_{\text{metals}}$, retaining only points for which $n_{\text{active}} = n_{\text{metals}}$.
This filter is applied consistently at every subsequent ranking and analysis step performed on the screening results.

Among the genuine non-equimolar compositions, the lowest, most favorable values of $T^*$ are concentrated in the binary and ternary Ta--Zr-containing subsets, with the Ta$_{0.5}$Zr$_{0.5}$ binary carbide showing the most negative $T^*$ of all sampled compositions, indicating that it is energetically stable without any mixing entropy contribution.
When binary and ternary compositions containing both Ta and Zr are excluded from consideration, the most stable genuine compositions instead concentrate around the Hf--Ta, Hf--Zr, and Nb--Ta binary subsets.
The best mixing energy per atom achieved within each subset improves modestly with increasing number of metals, from $-0.0247$~eV/atom for the binary subsets to $-0.0330$~eV/atom for the senary subset, while the number of compositions classified as synthesizable, with $T^*<3000$~K, increases substantially with subset size, reflecting the larger number of compositions sampled at higher metal counts.
\begin{table}[t]
\centering
\caption{Top-10 most synthesizable genuine non-equimolar compositions from the full screening (degenerate/collapsed-subset points excluded per the $n_{\text{active}}=n_{\text{metals}}$ filter).}
\label{tab:top10}
\resizebox{\columnwidth}{!}{%
\begin{tabular}{clcccccccc}
\hline\hline
Rank & Metals & $x_\mathrm{Hf}$ & $x_\mathrm{Nb}$ & $x_\mathrm{Ta}$ & $x_\mathrm{Ti}$ & $x_\mathrm{V}$ & $x_\mathrm{Zr}$ & $E_{\mathrm{mix}}$, eV/atom & $T^*$, K \\
\hline
1  & Ta, Zr        & 0.000 & 0.000 & 0.500 & 0.000 & 0.000 & 0.500 & $-0.025$ & $-826.8$ \\
2  & Ta, Zr        & 0.000 & 0.000 & 0.375 & 0.000 & 0.000 & 0.625 & $-0.023$ & $-801.0$ \\
3  & Ta, Zr        & 0.000 & 0.000 & 0.625 & 0.000 & 0.000 & 0.375 & $-0.021$ & $-745.9$ \\
4  & Hf, Ta, Zr     & 0.125 & 0.000 & 0.375 & 0.000 & 0.000 & 0.500 & $-0.029$ & $-701.0$ \\
5  & Ta, Zr        & 0.000 & 0.000 & 0.250 & 0.000 & 0.000 & 0.750 & $-0.017$ & $-691.3$ \\
6  & Hf, Ta, Zr     & 0.125 & 0.000 & 0.500 & 0.000 & 0.000 & 0.375 & $-0.028$ & $-676.3$ \\
7  & Hf, Ta, Zr     & 0.250 & 0.000 & 0.375 & 0.000 & 0.000 & 0.375 & $-0.031$ & $-672.9$ \\
8  & Ta, Zr        & 0.000 & 0.000 & 0.750 & 0.000 & 0.000 & 0.250 & $-0.016$ & $-664.4$ \\
9  & Hf, Ta, Zr     & 0.250 & 0.000 & 0.125 & 0.000 & 0.000 & 0.625 & $-0.026$ & $-664.1$ \\
10 & Hf, Ta, Zr     & 0.125 & 0.000 & 0.625 & 0.000 & 0.000 & 0.250 & $-0.026$ & $-663.8$ \\
\hline\hline
\end{tabular}}
\end{table}

\begin{figure}[H]
\centering
\includegraphics[width=\columnwidth]{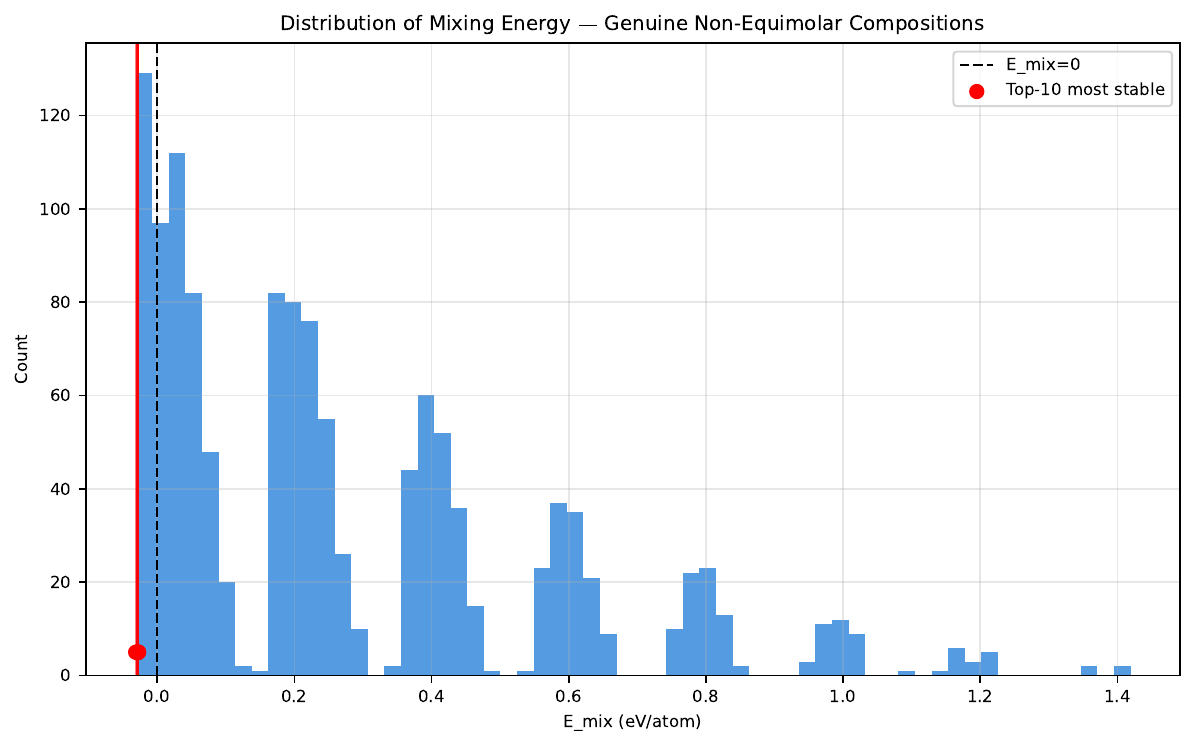}
\caption{Distribution of mixing energy $E_{\mathrm{mix}}$ across the 1{,}281 genuine non-equimolar compositions from the full screening. Red markers denote the ten most stable compositions listed in Table~\ref{tab:top10}.}
\label{fig:emix_dist}
\end{figure}

\subsubsection{PCA and t-SNE analysis of results}
To visualize the structure of the non-equimolar composition space screened in the preceding section and to assess how the synthesizability descriptor $T^*$ varies across it, the combined set of screening results and equimolar ensemble results is projected into two dimensions using both principal component analysis (PCA) and t-distributed stochastic neighbor embedding (t-SNE). The feature vector used for each composition point consists of the metal fractions $x_i$ for each of the six transition metals together with the mixing energy per atom, $E^{\text{atom}}_{\text{mix}}$, and the effective stabilization temperature, $T^*$. Features are standardized prior to dimensionality reduction.

\begin{figure*}[t]
\centering
\includegraphics[width=0.85\textwidth]{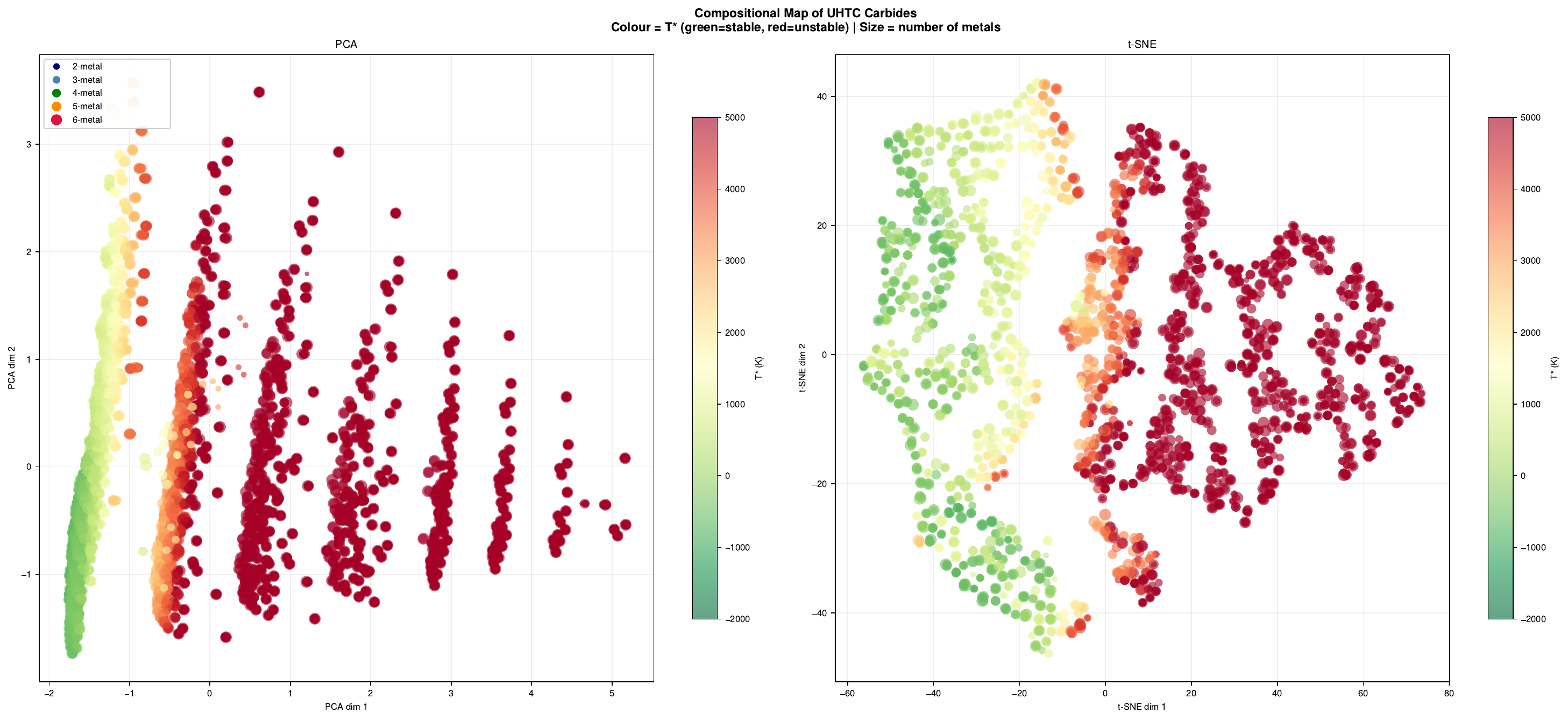}
\caption{PCA (left) and t-SNE (right) projections of the combined equimolar and non-equimolar screening results, colored by the effective stabilization temperature $T^*$ and sized by the number of active metals.}
\label{fig:compositional_map}
\end{figure*}

Figure~\ref{fig:compositional_map} shows the resulting PCA and t-SNE projections of the combined dataset, with each point colored by its $T^*$ value and sized according to the number of active metals in the composition. In both projections, compositions separate into distinct bands ordered primarily by the number of metals present, with two- and three-metal compositions clustering toward one side of the projected space and the senary composition occupying the opposite side. Within each band, a clear gradient in $T^*$ is visible, running from green, indicating compositions that are stable or easily synthesizable, to red, indicating compositions with large positive $T^*$ values that are predicted to be difficult to synthesize or likely unstable.

\section{Conclusions}
In conclusion, modern large atomistic model, namely MACE interatomic potential finely-tuned based on \textit{ab initio}  dataset of high-entropy carbides (HECs) allows us to perform high-throughput screening among equimolar and non-equimolar  HECs indicating those with highest probability to be synthesized.
Developed description and dataset of HEC structures opens a way to screening among large configurational space of non-equimolar HECs. 

\onecolumngrid

\section*{Supplementary Information}
\subsection*{Table S1: Top-100 Most Synthesizable Genuine Non-Equimolar Compositions}

\renewcommand{\thetable}{S1}

\begin{table}[p]
\centering
\footnotesize
\setlength{\tabcolsep}{4pt}
\caption{Top-100 most synthesizable genuine non-equimolar compositions from the full screening (7{,}581 raw points; degenerate/collapsed-subset points excluded via the $n_{\text{active}}=n_{\text{metals}}$ filter, leaving 1{,}281 genuine points), ranked by ascending $T^*$. Rows 1--50 of 100; continued in the following table.}
\label{tab:top100a}
\begin{tabular}{clcccccccc}
\hline\hline
Rank & Metals & $x_\mathrm{Hf}$ & $x_\mathrm{Nb}$ & $x_\mathrm{Ta}$ & $x_\mathrm{Ti}$ & $x_\mathrm{V}$ & $x_\mathrm{Zr}$ & $E_{\mathrm{mix}}$ (eV/atom) & $T^*$ (K) \\
\hline
1  & Ta,Zr        & 0.000 & 0.000 & 0.500 & 0.000 & 0.000 & 0.500 & $-0.025$ & $-826.809$ \\
2  & Ta,Zr        & 0.000 & 0.000 & 0.375 & 0.000 & 0.000 & 0.625 & $-0.023$ & $-801.049$ \\
3  & Ta,Zr        & 0.000 & 0.000 & 0.625 & 0.000 & 0.000 & 0.375 & $-0.021$ & $-745.856$ \\
4  & Hf,Ta,Zr     & 0.125 & 0.000 & 0.375 & 0.000 & 0.000 & 0.500 & $-0.029$ & $-700.981$ \\
5  & Ta,Zr        & 0.000 & 0.000 & 0.250 & 0.000 & 0.000 & 0.750 & $-0.017$ & $-691.323$ \\
6  & Hf,Ta,Zr     & 0.125 & 0.000 & 0.500 & 0.000 & 0.000 & 0.375 & $-0.028$ & $-676.334$ \\
7  & Hf,Ta,Zr     & 0.250 & 0.000 & 0.375 & 0.000 & 0.000 & 0.375 & $-0.031$ & $-672.859$ \\
8  & Ta,Zr        & 0.000 & 0.000 & 0.750 & 0.000 & 0.000 & 0.250 & $-0.016$ & $-664.427$ \\
9  & Hf,Ta,Zr     & 0.250 & 0.000 & 0.125 & 0.000 & 0.000 & 0.625 & $-0.026$ & $-664.135$ \\
10 & Hf,Ta,Zr     & 0.125 & 0.000 & 0.625 & 0.000 & 0.000 & 0.250 & $-0.026$ & $-663.755$ \\
11 & Hf,Ta,Zr     & 0.250 & 0.000 & 0.250 & 0.000 & 0.000 & 0.500 & $-0.030$ & $-661.969$ \\
12 & Hf,Ta,Zr     & 0.125 & 0.000 & 0.250 & 0.000 & 0.000 & 0.625 & $-0.024$ & $-625.942$ \\
13 & Hf,Ta,Zr     & 0.375 & 0.000 & 0.125 & 0.000 & 0.000 & 0.500 & $-0.026$ & $-614.224$ \\
14 & Hf,Ta,Zr     & 0.375 & 0.000 & 0.500 & 0.000 & 0.000 & 0.125 & $-0.026$ & $-613.838$ \\
15 & Hf,Ta,Zr     & 0.250 & 0.000 & 0.500 & 0.000 & 0.000 & 0.250 & $-0.027$ & $-602.410$ \\
16 & Hf,Ta,Zr     & 0.375 & 0.000 & 0.250 & 0.000 & 0.000 & 0.375 & $-0.028$ & $-593.996$ \\
17 & Hf,Ta,Zr     & 0.125 & 0.000 & 0.125 & 0.000 & 0.000 & 0.750 & $-0.019$ & $-590.946$ \\
18 & Hf,Ta,Zr     & 0.375 & 0.000 & 0.375 & 0.000 & 0.000 & 0.250 & $-0.027$ & $-584.027$ \\
19 & Hf,Ta         & 0.500 & 0.000 & 0.500 & 0.000 & 0.000 & 0.000 & $-0.017$ & $-569.021$ \\
20 & Hf,Zr         & 0.500 & 0.000 & 0.000 & 0.000 & 0.000 & 0.500 & $-0.017$ & $-564.868$ \\
21 & Hf,Ta,Zr     & 0.250 & 0.000 & 0.625 & 0.000 & 0.000 & 0.125 & $-0.022$ & $-561.599$ \\
22 & Hf,Ta,Zr     & 0.500 & 0.000 & 0.375 & 0.000 & 0.000 & 0.125 & $-0.024$ & $-560.076$ \\
23 & Hf,Ta,Zr     & 0.500 & 0.000 & 0.125 & 0.000 & 0.000 & 0.375 & $-0.023$ & $-553.863$ \\
24 & Hf,Zr         & 0.250 & 0.000 & 0.000 & 0.000 & 0.000 & 0.750 & $-0.013$ & $-528.304$ \\
25 & Hf,Nb,Ta,Zr  & 0.125 & 0.125 & 0.250 & 0.000 & 0.000 & 0.500 & $-0.027$ & $-524.708$ \\
26 & Hf,Nb,Ta,Zr  & 0.250 & 0.125 & 0.375 & 0.000 & 0.000 & 0.250 & $-0.030$ & $-518.521$ \\
27 & Hf,Ta,Zr     & 0.500 & 0.000 & 0.250 & 0.000 & 0.000 & 0.250 & $-0.023$ & $-517.594$ \\
28 & Nb,Ta         & 0.000 & 0.625 & 0.375 & 0.000 & 0.000 & 0.000 & $-0.015$ & $-511.566$ \\
29 & Nb,Ta         & 0.000 & 0.875 & 0.125 & 0.000 & 0.000 & 0.000 & $-0.008$ & $-510.224$ \\
30 & Hf,Nb,Ta,Zr  & 0.250 & 0.125 & 0.250 & 0.000 & 0.000 & 0.375 & $-0.029$ & $-509.045$ \\
31 & Nb,Ta         & 0.000 & 0.500 & 0.500 & 0.000 & 0.000 & 0.000 & $-0.015$ & $-496.611$ \\
32 & Ta,Zr         & 0.000 & 0.000 & 0.125 & 0.000 & 0.000 & 0.875 & $-0.008$ & $-493.407$ \\
33 & Nb,Ta         & 0.000 & 0.750 & 0.250 & 0.000 & 0.000 & 0.000 & $-0.012$ & $-486.748$ \\
34 & Hf,Zr         & 0.125 & 0.000 & 0.000 & 0.000 & 0.000 & 0.875 & $-0.008$ & $-483.423$ \\
35 & Hf,Nb,Ta,Zr  & 0.250 & 0.125 & 0.125 & 0.000 & 0.000 & 0.500 & $-0.025$ & $-482.617$ \\
36 & Hf,Nb,Ta,Zr  & 0.125 & 0.125 & 0.500 & 0.000 & 0.000 & 0.250 & $-0.025$ & $-481.045$ \\
37 & Hf,Nb,Ta,Zr  & 0.125 & 0.125 & 0.125 & 0.000 & 0.000 & 0.625 & $-0.022$ & $-480.820$ \\
38 & Nb,Ta,Zr     & 0.000 & 0.125 & 0.500 & 0.000 & 0.000 & 0.375 & $-0.020$ & $-470.203$ \\
39 & Hf,Nb,Ta,Zr  & 0.375 & 0.125 & 0.250 & 0.000 & 0.000 & 0.250 & $-0.027$ & $-468.434$ \\
40 & Hf,Nb,Ta,Zr  & 0.125 & 0.250 & 0.375 & 0.000 & 0.000 & 0.250 & $-0.027$ & $-466.297$ \\
41 & Nb,Ta,Zr     & 0.000 & 0.125 & 0.625 & 0.000 & 0.000 & 0.250 & $-0.018$ & $-465.997$ \\
42 & Nb,Ta         & 0.000 & 0.375 & 0.625 & 0.000 & 0.000 & 0.000 & $-0.013$ & $-464.916$ \\
43 & Hf,Zr         & 0.375 & 0.000 & 0.000 & 0.000 & 0.000 & 0.625 & $-0.013$ & $-462.379$ \\
44 & Hf,Ta,Zr     & 0.125 & 0.000 & 0.750 & 0.000 & 0.000 & 0.125 & $-0.014$ & $-456.259$ \\
45 & Hf,Nb,Zr     & 0.375 & 0.125 & 0.000 & 0.000 & 0.000 & 0.500 & $-0.019$ & $-455.077$ \\
46 & Hf,Nb,Ta,Zr  & 0.250 & 0.125 & 0.500 & 0.000 & 0.000 & 0.125 & $-0.024$ & $-453.167$ \\
47 & Hf,Zr         & 0.625 & 0.000 & 0.000 & 0.000 & 0.000 & 0.375 & $-0.013$ & $-451.146$ \\
48 & Nb,Ta,Zr     & 0.000 & 0.125 & 0.250 & 0.000 & 0.000 & 0.625 & $-0.017$ & $-448.098$ \\
49 & Hf,Nb,Ta,Zr  & 0.375 & 0.125 & 0.375 & 0.000 & 0.000 & 0.125 & $-0.024$ & $-447.528$ \\
50 & Hf,Nb,Ta,Zr  & 0.375 & 0.125 & 0.125 & 0.000 & 0.000 & 0.375 & $-0.024$ & $-445.272$ \\
\hline\hline
\end{tabular}
\end{table}

\begin{table}[p]
\centering
\footnotesize
\setlength{\tabcolsep}{4pt}
\caption{Top-100 most synthesizable genuine non-equimolar compositions (continued). Rows 51--100 of 100.}
\label{tab:top100b}
\begin{tabular}{clcccccccc}
\hline\hline
Rank & Metals & $x_\mathrm{Hf}$ & $x_\mathrm{Nb}$ & $x_\mathrm{Ta}$ & $x_\mathrm{Ti}$ & $x_\mathrm{V}$ & $x_\mathrm{Zr}$ & $E_{\mathrm{mix}}$ (eV/atom) & $T^*$ (K) \\
\hline
51 & Hf,Ta         & 0.375 & 0.000 & 0.625 & 0.000 & 0.000 & 0.000 & $-0.013$ & $-443.680$ \\
52 & Hf,Nb,Zr     & 0.250 & 0.125 & 0.000 & 0.000 & 0.000 & 0.625 & $-0.017$ & $-438.807$ \\
53 & Hf,Nb,Zr     & 0.250 & 0.250 & 0.000 & 0.000 & 0.000 & 0.500 & $-0.020$ & $-437.822$ \\
54 & Hf,Ta,Zr     & 0.625 & 0.000 & 0.125 & 0.000 & 0.000 & 0.250 & $-0.017$ & $-432.592$ \\
55 & Nb,Ta,Zr     & 0.000 & 0.250 & 0.500 & 0.000 & 0.000 & 0.250 & $-0.019$ & $-429.389$ \\
56 & Nb,Ta,Zr     & 0.000 & 0.125 & 0.375 & 0.000 & 0.000 & 0.500 & $-0.018$ & $-429.180$ \\
57 & Nb,Ta,Zr     & 0.000 & 0.500 & 0.375 & 0.000 & 0.000 & 0.125 & $-0.018$ & $-429.103$ \\
58 & Nb,Ta,Zr     & 0.000 & 0.250 & 0.625 & 0.000 & 0.000 & 0.125 & $-0.017$ & $-427.011$ \\
59 & Nb,Ta,Zr     & 0.000 & 0.375 & 0.375 & 0.000 & 0.000 & 0.250 & $-0.020$ & $-426.386$ \\
60 & Hf,Nb,Zr     & 0.500 & 0.125 & 0.000 & 0.000 & 0.000 & 0.375 & $-0.018$ & $-425.023$ \\
61 & Hf,Nb,Ta,Zr  & 0.250 & 0.250 & 0.250 & 0.000 & 0.000 & 0.250 & $-0.025$ & $-422.088$ \\
62 & Hf,Nb,Ta,Zr  & 0.125 & 0.125 & 0.375 & 0.000 & 0.000 & 0.375 & $-0.023$ & $-420.009$ \\
63 & Hf,Nb,Ta,Zr  & 0.125 & 0.250 & 0.250 & 0.000 & 0.000 & 0.375 & $-0.024$ & $-416.025$ \\
64 & Nb,Ta,Zr     & 0.000 & 0.625 & 0.250 & 0.000 & 0.000 & 0.125 & $-0.016$ & $-415.091$ \\
65 & Nb,Ta         & 0.000 & 0.250 & 0.750 & 0.000 & 0.000 & 0.000 & $-0.010$ & $-407.915$ \\
66 & Hf,Nb,Ta,Zr  & 0.125 & 0.375 & 0.250 & 0.000 & 0.000 & 0.250 & $-0.023$ & $-406.360$ \\
67 & Nb,Ta,Zr     & 0.000 & 0.250 & 0.375 & 0.000 & 0.000 & 0.375 & $-0.019$ & $-405.599$ \\
68 & Nb,Ta,Zr     & 0.000 & 0.500 & 0.250 & 0.000 & 0.000 & 0.250 & $-0.018$ & $-397.721$ \\
69 & Hf,Nb,Ta,Zr  & 0.250 & 0.250 & 0.375 & 0.000 & 0.000 & 0.125 & $-0.023$ & $-396.284$ \\
70 & Hf,Nb,Ta,Zr  & 0.125 & 0.125 & 0.625 & 0.000 & 0.000 & 0.125 & $-0.018$ & $-394.533$ \\
71 & Hf,Nb,Ta,Zr  & 0.250 & 0.250 & 0.125 & 0.000 & 0.000 & 0.375 & $-0.022$ & $-393.219$ \\
72 & Nb,Ta,Zr     & 0.000 & 0.125 & 0.750 & 0.000 & 0.000 & 0.125 & $-0.012$ & $-391.280$ \\
73 & Hf,Nb,Ta,Zr  & 0.125 & 0.500 & 0.125 & 0.000 & 0.000 & 0.250 & $-0.020$ & $-391.167$ \\
74 & Hf,Nb,Ta,Zr  & 0.125 & 0.250 & 0.500 & 0.000 & 0.000 & 0.125 & $-0.020$ & $-388.497$ \\
75 & Nb,Ta,Zr     & 0.000 & 0.375 & 0.500 & 0.000 & 0.000 & 0.125 & $-0.016$ & $-387.094$ \\
76 & Hf,Ta         & 0.250 & 0.000 & 0.750 & 0.000 & 0.000 & 0.000 & $-0.009$ & $-384.067$ \\
77 & Hf,Ta,Zr     & 0.625 & 0.000 & 0.250 & 0.000 & 0.000 & 0.125 & $-0.015$ & $-382.758$ \\
78 & Hf,Nb,Zr     & 0.125 & 0.250 & 0.000 & 0.000 & 0.000 & 0.625 & $-0.015$ & $-382.719$ \\
79 & Nb,Ta,Zr     & 0.000 & 0.250 & 0.250 & 0.000 & 0.000 & 0.500 & $-0.017$ & $-381.439$ \\
80 & Nb,Ta,Zr     & 0.000 & 0.750 & 0.125 & 0.000 & 0.000 & 0.125 & $-0.012$ & $-378.430$ \\
81 & Hf,Nb,Ta,Zr  & 0.125 & 0.375 & 0.125 & 0.000 & 0.000 & 0.375 & $-0.020$ & $-375.194$ \\
82 & Hf,Ta         & 0.625 & 0.000 & 0.375 & 0.000 & 0.000 & 0.000 & $-0.011$ & $-371.368$ \\
83 & Hf,Nb,Zr     & 0.125 & 0.125 & 0.000 & 0.000 & 0.000 & 0.750 & $-0.012$ & $-367.295$ \\
84 & Hf,Nb,Ta,Zr  & 0.500 & 0.125 & 0.125 & 0.000 & 0.000 & 0.250 & $-0.019$ & $-363.675$ \\
85 & Hf,Nb,Ta     & 0.125 & 0.500 & 0.375 & 0.000 & 0.000 & 0.000 & $-0.015$ & $-362.159$ \\
86 & Hf,Nb,Ta     & 0.125 & 0.250 & 0.625 & 0.000 & 0.000 & 0.000 & $-0.014$ & $-362.058$ \\
87 & Hf,Nb,Ta,Zr  & 0.125 & 0.250 & 0.125 & 0.000 & 0.000 & 0.500 & $-0.019$ & $-359.084$ \\
88 & Hf,Nb,Ta,Zr  & 0.375 & 0.250 & 0.125 & 0.000 & 0.000 & 0.250 & $-0.020$ & $-354.484$ \\
89 & Hf,Nb,Ta,Zr  & 0.250 & 0.375 & 0.250 & 0.000 & 0.000 & 0.125 & $-0.020$ & $-353.333$ \\
90 & Hf,Nb,Zr     & 0.375 & 0.250 & 0.000 & 0.000 & 0.000 & 0.375 & $-0.016$ & $-351.529$ \\
91 & Nb,Ta,Zr     & 0.000 & 0.250 & 0.125 & 0.000 & 0.000 & 0.625 & $-0.013$ & $-347.429$ \\
92 & Hf,Nb,Ta     & 0.125 & 0.375 & 0.500 & 0.000 & 0.000 & 0.000 & $-0.015$ & $-347.380$ \\
93 & Hf,Nb,Ta,Zr  & 0.375 & 0.250 & 0.250 & 0.000 & 0.000 & 0.125 & $-0.020$ & $-344.574$ \\
94 & Nb,Ta,Zr     & 0.000 & 0.500 & 0.125 & 0.000 & 0.000 & 0.375 & $-0.014$ & $-341.834$ \\
95 & Hf,Nb,Ta,Zr  & 0.125 & 0.375 & 0.375 & 0.000 & 0.000 & 0.125 & $-0.018$ & $-340.180$ \\
96 & Hf,Nb,Ta,Zr  & 0.125 & 0.500 & 0.250 & 0.000 & 0.000 & 0.125 & $-0.018$ & $-339.367$ \\
97 & Hf,Nb,Ta     & 0.125 & 0.625 & 0.250 & 0.000 & 0.000 & 0.000 & $-0.013$ & $-336.433$ \\
98 & Hf,Nb,Ta     & 0.375 & 0.250 & 0.375 & 0.000 & 0.000 & 0.000 & $-0.015$ & $-332.398$ \\
99 & Hf,Nb,Ta,Zr  & 0.250 & 0.375 & 0.125 & 0.000 & 0.000 & 0.250 & $-0.019$ & $-332.178$ \\
100 & Hf,Nb,Zr    & 0.250 & 0.375 & 0.000 & 0.000 & 0.000 & 0.375 & $-0.015$ & $-330.110$ \\
\hline\hline
\end{tabular}
\end{table}

\twocolumngrid
\clearpage
\bibliography{references}

\end{document}